\documentclass[lettersize,journal]{IEEEtran}
\usepackage{amsmath,amsfonts}
\usepackage{algorithmic}
\usepackage{bm}
\usepackage{algorithm}
\usepackage{array}
\usepackage[caption=false,font=normalsize,labelfont=sf,textfont=sf]{subfig}
\usepackage{textcomp}
\usepackage{stfloats}
\usepackage{url}
\usepackage{verbatim}
\usepackage{graphicx}
\usepackage{cite}
\usepackage{multirow}
\begin{document}

\title{CvFormer: Cross-view transFormers with Pre-training for fMRI Analysis of Human Brain}

\author{Xiangzhu Meng$^*$, Qiang Liu, IEEE, Member, Shu Wu, IEEE, Senior Member, Liang Wang, IEEE, Fellow \thanks{Xiangzhu Meng and Wei Wei are corresponding authors (denoted by $\ast$)} \thanks{X. Meng is with the Center for Research on Intelligent Perception and Computing, Institute of Automation, Chinese Academy of Sciences, Beijing, China (xiangzhu.meng@cripac.ia.ac.cn)}  \thanks{Q. Liu, W. Shu and L. Wang are with Center for Research on Intelligent Perception and Computing Institute of Automation, Chinese Academy of Sciences, and School of Artificial Intelligence, University of Chinese Academy of Sciences (qiang.liu, shu.wu, liang.wang\}@nlpr.ia.ac.cn)} }

% \author{IEEE Publication Technology,~\IEEEmembership{Staff,~IEEE,}
%         % <-this % stops a space
% \thanks{This paper was produced by the IEEE Publication Technology Group. They are in Piscataway, NJ.}% <-this % stops a space
% \thanks{Manuscript received April 19, 2021; revised August 16, 2021.}}

% The paper headers
\markboth{Journal of \LaTeX\ Class Files,~Vol.~14, No.~8, August~2021}%
{Shell \MakeLowercase{\textit{et al.}}: CvFormer: Cross-view transFormers with Pre-training for fMRI Analysis of Human Brain}

\IEEEpubid{\copyright~2021}
% Remember, if you use this you must call \IEEEpubidadjcol in the second
% column for its text to clear the IEEEpubid mark.

\maketitle

\begin{abstract}
In recent years, functional magnetic resonance imaging (fMRI) has been widely utilized to diagnose neurological disease, by exploiting the region of interest (RoI) nodes as well as their connectivities in human brain. However, most of existing works only rely on either RoIs or connectivities, neglecting the potential for complementary information between them. To address this issue, we study how to discover the rich cross-view information in fMRI data of human brain. This paper presents a novel method for cross-view analysis of fMRI data of the human brain, called Cross-view transFormers (CvFormer). CvFormer employs RoI and connectivity encoder modules to generate two separate views of the human brain, represented as RoI and sub-connectivity tokens. Then, basic transformer modules can be used to process the RoI and sub-connectivity tokens, and cross-view modules integrate the complement information across two views.
Furthermore, CvFormer uses a global token for each branch as a query to exchange information with other branches in cross-view modules, which only requires linear time for both computational and memory complexity instead of quadratic time. 
To enhance the robustness of the proposed CvFormer, we propose a two-stage strategy to train its parameters. To be specific, RoI and connectivity views can be firstly utilized as self-supervised information to pre-train the CvFormer by combining it with contrastive learning and  then fused to finetune the CvFormer using label information. Experiment results on two public ABIDE and ADNI datasets can show clear improvements by the proposed CvFormer, which can validate its effectiveness and superiority.

\end{abstract}


\section{Introduction}
With the rapid development of modern medical era, magnetic resonance imaging (MRI) \cite{terreno2010challenges} technologies have been proven to be a valuable tool in the investigation of neurological issues, particularly in the diagnosis of neurological disorders. In particular, functional magnetic resonance imaging (fMRI) \cite{bennett2010reliable} is one of the non-invasive techniques to temporal dynamics of blood oxygen level dependency (BOLD) \cite{glover2011overview} response. In recent years, fMRI data of human brain has been widely exploited to understand functional activities and organization of human brain. Recent research \cite{thiebaut2022emergent} suggests that functional connectivities \cite{lee2022solving} among brain regions of interest (RoI) are crucial in determining behavior, cognition, and brain dysfunction.

In recent years, machine learning methods \cite{larranaga2006machine} have been widely used to build the corresponding diagnosis model. Among these works, shadow learning-based methods usually follow two stages, where brain networks are first processed to produce an embedding of the human brain, and then a classical classifier is utilized to divide the learned data into corresponding groups. However, these methods are prone to inducing substantial errors in the second stage if the brain features from the first stage are not reliable. With the established power of deep learning methods \cite{salehinejad2017recent}, most works based on Convolution Neural Networks (CNN) \cite{li2021survey} and Graph Neural Networks (GNN) \cite{zhou2020graph} have been widely developed to extract spatial, temporal and connective patterns of fMRI time series for brain disorder diagnose. For instance, BrainNetCNN \cite{kawahara2017brainnetcnn} leverages the topological locality of brain networks to predict cognitive and motor developmental outcome scores for infants born preterm; BrainGNN \cite{li2021braingnn} designs novel RoI-aware graph convolutional layers that leverage the topological and functional information of fMRI. Moreover, transformers \cite{vaswani2017attention} have been studied over different types of data, and there also exist several transform-based works for human brain, such as BRAINNETTF \cite{kan2022brain}, which leverages the unique properties of brain network data to maximize the power of transformer-based models for brain network analysis. Unfortunately, limited brain data might result in erratic results of the above diagnosis models due to expensive costs of data acquisition.

\subsection{Motivations}
%For many real-world applications, dimension reduction is a necessary tool because the dimensions of samples are higher and higher. As an important family of DR, sparse subspace learning methods have attracted wide attentions due to their excellent performances. With the wide applications of multi-view data, a multiview SSL method is necessary to fill the blank of this field. Although some multi-view methods are proposed, the following 2 aspects have not been investigated comprehensively and thoroughly:

%they mainly focus on RoIs or connectivity information in human brain. This manner fails to mine the rich and complementary information between two views of RoIs and connectivity in human brain. Moreover, transformer-based models have been studied over different types of data, yet there exist less transform-based works for human brain. 

Even though most existing methods have obtained promising performance for brain disorder diagnosis in certain situations, there still remain two important aspects that have yet to be thoroughly investigated comprehensively.
\begin{itemize}
    \item[1.] Different from single-view data, cross-view data \cite{meng2022unified} can reflect different properties of human brain, i.e. node-level and edge-level information of human brain. However, most existing methods mainly focus on RoIs or connectivity information in human brain, which limits their ability to fully exploit the diverse and complementary information available across multiple views.
    
    \item[2.] There exist less transform-based works for fMRI analysis of human brain. The main reason is that large-scale brain data can be difficult to collect to train such models. Thus, it's necessary to provide a suitable manner to pre-train the transformer-based models when facing limited human brain data.

\end{itemize}

\subsection{Contributions}
This paper proposes a cross-view transformers (CvFormer) to jointly model the human brain from two perspectives, namely regions of interest (RoIs) and their connectivity. CvFormer employs RoI and connectivity encoder modules to generate RoI and sub-connectivity tokens as two views. These tokens are processed through separate branches based on transformer modules. Then, RoI and sub-connectivity tokens can be purely fused by cross-view modules multiple times to complement each other. 
% Finally, we can combine cross-view global tokens as the ultimate brain representation. 
Notably, CvFormer uses a global token for each branch as a query to exchange information based on cross-view modules in an efficient manner, reducing the computational and memory complexity from a quadratic to a linear time complexity. 
Finally, we propose a two-stage strategy to train its parameters. We combine RoI and connectivity views with contrastive learning to pre-train the CvFormer and fuse cross-view information to finetune the CvFormer using label information.
We evaluate the effectiveness of the proposed CvFormer on two neurological disorders datasets of ADNI and ABIDE. The major contributions of this paper can be summarized as follows:
\begin{itemize}
    \item[1.] CvFormer can simultaneously consider the diversity and complementary information between two views of human brain, leading to more distinguishable and informative brain representations. 
    
    \item[2.] A two-stage strategy is proposed to train the parameters of CvFormer, where RoI and connectivity views can be combined with contrastive learning to pre-train the CvFormer and fused to finetune the CvFormer.

    \item[3.] Massive experimental results on ADNI and ABIDE datasets can validate the effectiveness and superiority of the proposed CvFormer, corresponding to Alzheimer's Disorder (AD) and Autism Spectrum Disorder (ASD).
\end{itemize}

\begin{figure*}
\centering
\includegraphics[width=0.75\textwidth]{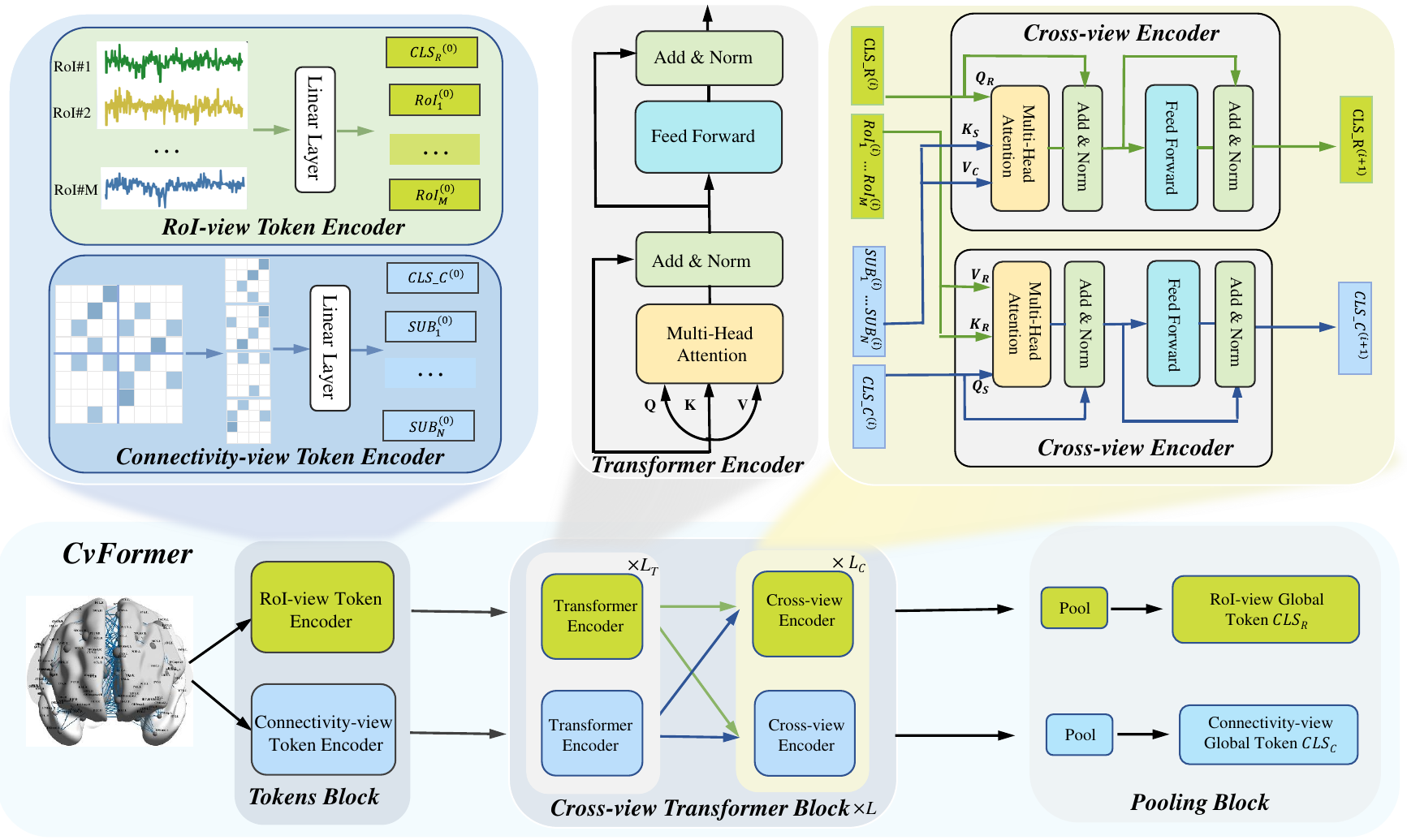}
\caption{Flowchart of CvFormer framework, consisting of three following parts, tokens block, cross-view transformer block, and pooling block. These blocks serve the purpose of generating cross-view representations, leveraging the complementary information between views, and ultimately pooling the global cross-view information for human brain representation, respectively.} \label{fig:architecture}
\end{figure*}

\section{Method} 
\subsection{Overall Architecture}
In this section, we introduced a novel cross-view transformers (CvFormer) model for brain disorder diagnosis, as shown in Fig. \ref{fig:architecture}. The CvFormer model comprises three key components: (1) Tokens block contains two branches, i.e. RoI-view and Connecivity-view token encoders, which generate the initial RoI-view and Connecivity-view tokens of human brains. (2) Cross-view transformer block contains two branches based on transformer encoder to process RoI-view and connectivity-view tokens, and leverages cross-view encoder to learn complementary information between views. (3) Pooling block is utilized to extract the global cross-view representations, corresponding to RoI and connecivity views.

\subsection{Cross-View Transformer}
Cross-view transformer mainly consists of three following blocks.

\textbf{Tokens block}
is utilized to transform preprocessed fMRI data of human brain into cross-view tokens, consisting of RoI-view and connectivity-view token encoders.
For RoI-view token encoder, given $M$ RoIs of human brain, the initial feature of each RoI usually can be represented by the time series or connectivity profile. Then, one linear layer is used to project the initial features to latent space. Adding it with positional embeddings, we can get RoI-view tokens as 
\begin{small}
\begin{equation}
        \left[\bm{CLS}_R^{(0)}, \bm{RoI}_1^{(0)}, \bm{RoI}_2^{(0)}, \cdots, \bm{RoI}_M^{(0)} \right],
\end{equation}
\end{small}
where $\bm{CLS}_R$ denotes the class token to exploit the global RoI-view
information. $\bm{RoI}_i$ is the token of $i$th RoI node of human brain.
For connectivity-view token encoder, we first convert a connectivity network into a sequence of $N$ patch tokens by dividing it with a certain patch size, where each patch token denotes the local connectivity pattern of human brain. Then, we linearly project each patch and add positional embeddings into connectivity-view tokens, which can be expressed as 
\begin{small}
\begin{equation}
       \left[\bm{CLS}_C^{(0)}, \bm{SUB}_1^{(0)}, \bm{SUB}_2^{(0)}, \cdots, \bm{SUB}_N^{(0)} \right],
\end{equation}    
\end{small}
where $\bm{CLS}_C$ denotes the class token to exploit the global RoI-view
information. $\bm{SUB}_i$ is the token of $i$th local connectivity pattern of human brain.

\textbf{Cross-view transformer block}
is the core component of CvFormer, containing two transfomer-based branches that encode rich information between RoI and connectivity views. Cross-view encoder is adopted in cross-view transformer block to enforce RoI and connectivity views to learn with each other.
 A transformer encoder is composed of a sequence of blocks where each block contains multi-headed self-attention (MSA) with a feed-forward network (FFN). For each head of MSA module, query matrix $\bm{Q}$, key matrix $\bm{K}$ and value matrix $\bm{V}$ are generated by linearly projecting RoI-view or connectivity tokens layers. The scaled dot-product self-attention is applied on $\bm{Q}$, $\bm{K}$ and $\bm{V}$ can be expressed as
\begin{equation}
       Attention(\bm{Q}, \bm{K}, \bm{V}) = softmax(\frac{\bm{Q} \bm{K}^T}{\sqrt{d_k}})\bm{V},
\end{equation}
where $d_k$ is the dimension of $\bm{Q}$ and $\bm{K}$. The softmax function is applied to each row of the attention score that reflects the relationship between $\bm{Q}$ and $\bm{K}$.
FFN contains two-layer linear layers with expanding ratio $r$ at the hidden layer, and one GELU non-linearity is applied after the first linear layer. Layer normalization (LN) is applied before every block, and residual shortcuts are applied after every block. For RoI-view tokens $\bm{X}_R^{(l)}=\left[\bm{CLS}_C^{(0)}, \bm{SUB}_1^{(0)}, \bm{SUB}_2^{(0)}, \cdots, \bm{SUB}_N^{(0)} \right]$, the processing of the $l$th transformer encoder of RoI-view branch can be expressed as follows:
\begin{equation}\label{eq:MSA}
   \bm{X}_R^{(l)} = LN(\bm{X}_R^{(l)}+MSA(\bm{X}_R^{(l)})),
\end{equation}
\begin{equation}\label{eq:FFN}
   \bm{X}_R^{(l+1)} = LN(\bm{X}_R^{(l)}+FFN(\bm{X}_R^{(l)})).
\end{equation}
Similarly, we can also get the processing of the $l$th transformer encoder of connectivity-view branch following by Eq.(\ref{eq:MSA})-Eq.(\ref{eq:FFN}).

 To integrate complement information between RoI and connectivity views, cross-view encoder is proposed to enhance the key information of human brain. Generally, the CLS token already learns global information among all tokens in such view, interacting with context tokens at the other branch helps to integrate cross-view information. Thus, the basic idea of cross-view encoder is to involve the CLS token of one branch and context tokens of the other branch. That is, we first utilize the CLS token at each branch as one query to exchange information among context tokens from the other branch by MSA. Then, FFN is applied after cross-view fusion. LN is applied before every block, and residual shortcuts after every block. For each head of MSA of cross-view encoder, its processing can be expressed as follows:
\begin{equation}
\small
   \bm{CLS}_R^{(l+1)} = Attention( \bm{CLS}_R^{(l)}, Content_{C}^{(l)}, Content_{C}^{(l)}),
\end{equation}
\begin{equation}
\small
   \bm{CLS}_C^{(l+1)} = Attention( \bm{CLS}_C^{(l)}, Content_{R}^{(l)},Content_{R}^{(l)}).
\end{equation}

where $Content_{C}^{(l)}=[\bm{SUB}_1^{(l)}, \cdots, \bm{SUB}_N^{(l)}]$ and $Content_{R}^{(l)}=[\bm{RoI}_1^{(l)}, \cdots, \bm{RoI}_M^{(l)}]$.
After fusing cross-view information, the CLS token interacts with its own patch tokens again at the next transformer encoder, where it is able to pass the learned information from the other branch to its own patch tokens, to enrich the representation of each patch token. Therefore, cross-view encoder can effectively preserve the interactive information between RoI and connectivity views. Besides, the computation and memory complexity of generating the attention map in cross-view encoder are linear rather than quadratic time.

\textbf{Pooling block} is proposed to obtain the global cross-view information via pooling operation after the processing of cross-view transformer blocks. To be specific, the pooling operation employs the CLS token as final embedding for each branch. That, we can get the global tokens $\bm{CLS}_R$ and $\bm{CLS}_C$ for ROI and connectivity views, as shown in Fig. \ref{fig:architecture}.
% We adopt multi-layer perception (MLP) to produce the probability scores of target labels for two views, and then fuse these scores as final output for brain disorder diagnosis. In training the whole CvFormer, we employ the cross entropy as loss function and stochastic gradient descent (SGD) method to update all weight parameters.

\subsection{Two-stage Training}
To enhance the robustness of the proposed CvFormer, we propose a two-stage training strategy to learn its parameters. For the first stage, RoI and connectivity views can be utilized as self-supervised information to pre-train the CvFormer by combining it with contrastive learning. For the second stage, we can fuse the cross-view information to finetune the CvFormer through refined label information. 

\subsubsection{Self-supervised Pre-training}
Through the processing of CvFormer for human brain, we can obtain global tokens $\bm{CLS}_R$ and $\bm{CLS}_C$ for ROI and connectivity views. For any instance of human brain, its embedding generated in ROI view (or connectivity view),  is treated as the anchor, the embedding of it generated in the other view,  forms the positive sample, and the other embeddings in the two views are naturally regarded as negative samples. In this way, we can employ a contrastive objective that distinguishes the embeddings of the same instance in these two different views from other instances' embeddings. Mirroring the InfoNCE objective in our self-supervised pre-training setting, we can define the pairwise objective for positive pairs  $\{ (\bm{u}_i, \bm{v}_i ) \}$ as
\begin{equation}
\bm{Loss}_{C L} = \sum_{i=1}^N { \frac{exp(\theta(\bm{u}_i, \bm{v}_i))/\tau}{exp(\theta(\bm{u}_i, \bm{v}_i))/\tau+neg(i)}}
\end{equation}
where $neg(i) = \sum_{k \neq i}{(exp(\theta(\bm{u}_i, \bm{v}_k))+exp(\theta(\bm{u}_k, \bm{v}_i)))/\tau}$, and $\tau$ is a temperature parameter. $\theta(\bm{u}_i, \bm{v}_i)=s(h(\bm{u}_i), h(\bm{u}_i))$, where $s(\cdot, \cdot)$ is the cosine similarity and $h(\cdot)$ is a multi-layer perception (MLP).

\subsubsection{Fine-tuning}
After the processing of cross-view transformer blocks, the CLS token is utilized as final embedding for each branch. We adopt the MLP to produce the probability scores of target labels for two views, and then fuse these scores as final output for brain disorder diagnosis. In fine-tuning the whole CvFormer, we combine the cross entropy and contrastive loss as the final loss function, which can be formulated as 
\begin{equation}\label{eq:loss}
\bm{Loss} = \bm{Loss}_{C E} + \lambda \bm{Loss}_{C L}
\end{equation}
where $\lambda>0$ is a hyper-parameter to balance the cross entropy and contrastive loss terms.
Finally, we utilize the stochastic gradient descent (SGD) method to update all weight parameters.

\section{Experiments}

\begin{table*}
\centering
\caption{Comparison with different baselines (\%).}
\label{tab:performance}
\begin{tabular}{c|ccccc|ccccc}
\hline
\multirow{2}{*}{Baselines} & \multicolumn{5}{|c|}{Accuracy} & \multicolumn{5}{c}{Recall}\\
\cline{2-6} \cline{7-11}
 & ADNI & UCLA & USM & NYU & UM & ADNI & UCLA & USM & NYU & UM \\
\hline
PCA & 64.98 & 76.34 & 64.84 & 64.84 & 82.62 & 54.71 & 75.73 & 66.94 & 63.44 & 79.10 \\
SVM & 56.95 & 72.18 & 48.42 & 54.92 & 68.37 & 54.57 & 72.13 & 48.48 & 53.76 & 65.26 \\
MLP & 80.90 & 86.77 & 80.45 & 83.92 & 87.07 & 75.99 & 83.71 & 68.08 & 81.13 & 81.52 \\
BrainGNN & 84.05 & 86.16 & 82.37 & 89.86 & 86.77 & 83.48 & 92.78 & 71.03 & 87.88 & 85.78 \\
BrainnetCNN & 84.20 & 87.56 & 81.19 & 89.02 & 88.72 & 83.24 & 92.22 & 70.23 & 89.17 & 86.25 \\
Graphnormer & 84.98 & 91.32 & 85.62 & 86.67 & 88.27 & 86.39 & 91.77 & 86.80 & 84.76 & 87.24 \\
BRAINNETTF & 85.99 & 89.63 & 81.49 & 92.14 & 94.25 & 86.29 & 88.87 & 83.78 & 91.54 & 92.68 \\
CvFormer & \textbf{89.09} & \textbf{93.29} & \textbf{87.51} & \textbf{92.60} & \textbf{95.96} & \textbf{87.35} & \textbf{92.42} & \textbf{87.84} & \textbf{93.26} & \textbf{95.92} \\
\hline
\end{tabular}
\end{table*}

\subsection{Experiment Setting}
\textbf{Datasets.} To validate the effectiveness of the proposed CvFormer, massive experiments are performed on two types of brain neuro-science disorder diagnosis. Specifically, Alzheimer's Disease Neuroimaging Initiative (ADNI)\footnote{http://www.adni-info.org.} contains the neuRoImaging data of 128 different patients, consisting of 54 Early Mild Cognitive Impairment (EMCI), 38 Late Mild Cognitive Impairment Impairment (LMCI), and 34 Alzheimer's Disease (AD). Autism Brain Imagine Data Exchange (ABIDE)\footnote{http://preprocessed-connectomes-project.org/abide/download.html} contains 1112 subjects, which are composed of structural and resting state fMRI data along with the corresponding phenotypic information.

\textbf{Preprocessing.} For ABIDE dataset, we downloaded the preprocessed rs-fMRI series data of the top four largest sites (UM, NYU, USM, UCLA) from the preprocessed ABIDE dataset with Configurable Pipeline for the Analysis of Connectomes (CPAC), band-pass filtering (0.01 - 0.1 Hz), no global signal regression, parcellatinng each brain into 90 RoIs by the Automated Anatomical Labeling (AAL) atlas. For ADNI dataset, we utilized the standard preprocessing procedure to process original fMRI data as work \cite{cui2022braingb}, and each human brain was also parcellated into 90 RoIs using AAL atlas. For both the ABIDE and ADNI datasets, the mean time series of the RoIs were used to compute the functional connectivity network (FCN) by computing the correlation matrix. The FCN was then represented as RoI-view features, where each row of the FCN represented a node in the human brain. A connectivity-view feature was obtained by drawing an edge between each pair of RoIs with a correlation larger than the 70-th percentile of the correlation values.

\textbf{Baselines.} We evaluated the effectiveness of the proposed CvFormer in classifying brain networks by comparing it with the following baselines, including Principle Component Analysis (PCA) \cite{wold1987principal}, Support Vector Machine (SVM) \cite{noble2006support}, Multilayer Perceptron (MLP), Graph Convolutional Networks (GCN) \cite{wu2019simplifying}, BrainGNN \cite{li2021braingnn}, BrainNetCNN \cite{kawahara2017brainnetcnn}, Graphnomer \cite{ying2021transformers}, and BRAINNETTF \cite{kan2022brain}.

%other methods. Two types of methods are chosen to compare their performance with our proposed framework. The former mainly focuses on the flatten vector of brain connectivity network, 

\textbf{Implement Details.} For all datasets, we randomly select 70\% of the samples as training samples, 10\% of samples as validating samples, and the remaining 20\% of samples as testing samples at each iteration. We repeatedly run the above validation process ten times for all methods and use the accuracy of classification as the evaluation index. The above computational experiments are performed on a server running Ubuntu Linux 18.04 64bit with 56 CPUs (Xeon E5-2660 v4) at 3.2 GHz and 8 GPUs (NVIDIA TITAN Xp). The experimental setting for the baselines is based on the methodology described in the corresponding original papers. Finally, the results of the experiments are aggregated and summarized.

\subsection{Evaluation Results and Analysis}

As illustrated in Table \ref{tab:performance}, we could find that the proposed CvFormer performs other methods in most situations. The main reason was that cross-view representation of human brain provides more comprehensive information as compared to single-view approaches (such as Region of Interest or connectivity-based methods). Additionally, the cross-view encoder effectively captured complementary information from both the Region of Interest and connectivity views, further enhancing the performance of CvFormer. Among single-view works, we observed that transformer-based methods achieved remarkable for brain disorder diagnosis, demonstrating the modeling power of transformers for human brains. In comparison to deep learning-based methods, traditional methods were unable to deliver comparable results, suggesting that deep learning-based techniques are more effective in automatically discovering meaningful information. However, these methods were limited by their reliance on single-view information, failing to fully utilize the rich information in the human brain. Therefore, it's necessary to first extract the prior knowledge for the subsequent neurological disorders diagnosis.

 % cross-view fusion modules and 

% \vspace{-1cm}
\subsection{Ablation Analysis}
We perform an extensive ablation study on ADNI dataset to evaluate the impact of individual components and to assess the effectiveness of our proposed CvFormer approach with varying architectural specifications.

\subsubsection{Efforts of Sub-modules}
We have conducted massive experiments on ADNI dataset to evaluate the impact of sub-modules of CvFormer. As demonstrated by the results presented in Table \ref{tab:abalation}, the proposed CvFormer exhibited superior performance when simultaneously utilizing these sub-modules. In comparison to utilizing single RoI or connectivity view, two views could result in more rich information than single view. Despite the use of transformer encoder to combine information between two views, it failed to achieve better performance without cross-view encoder. 
%In terms of fusion strategy, the performances of both the sum and concat strategies were found to be comparable. 
Consequently, the integration of both the RoI-view and connectivity-view branches could lead to the highest accuracy, indicating that these two branches learn complementary features.

\begin{table}
\centering
\caption{Comparison with different modules (\%).}
\label{tab:abalation}
\begin{tabular}{c|c|c|cc}
\hline
RoI & Connectivity & Cross-view encoder &  Accuracy & Recall \\
\hline
\checkmark & $\times$ & $\times$ &  81.75 & 76.24  \\
$\times$ & \checkmark & $\times$ & 80.97 & 77.96 \\
\checkmark & \checkmark& $\times$ & 82.36 & 78.14 \\
\checkmark & \checkmark & \checkmark & \textbf{86.74} & \textbf{82.57} \\
\hline
\end{tabular}
\end{table}

\subsubsection{Efforts of Pretraining}
We evaluate the impact of self-supervised pertraining for the proposed CvFormer on ADNI dataset. As demonstrated by the results presented in Table \ref{tab:pretrain}, CvFormer can obtain more promising performance when training it with self-supervised pertaining. The main reason is that training transformers-based models usually depends on large-scale brain data. In comparison to directly training the CvFormer, this manner can initialize more suitable parameters for limited human brain data. 
Thus, it's necessary to provide a suitable manner to pre-train the transformer-based models when facing limited human brain data.
Therefore, using self-supervised pretraining as a suitable manner to pre-train the transformer-based models is very necessary and valuable when facing limited human brain data.

\begin{table}
\centering
\caption{Comparison with(without) pretraining(\%).}
\label{tab:pretrain}
\setlength{\tabcolsep}{4mm}{
\begin{tabular}{c|cc}
\hline
Pretraining & Accuracy & Recall\\
\hline
\checkmark  &  \textbf{89.09} & \textbf{87.35} \\
$\times$  &  83.23 & 82.78 \\
\hline
\end{tabular}
}
\end{table}

\begin{table}[htbp]
\small
\caption{Comparison (\%)  with different $\lambda$.}
\label{tab:param}
\centering
\begin{tabular*}{0.45\textwidth}{@{\extracolsep{\fill}}c|ccccc}  
\hline
 $\lambda_1$ & 0.01 & 0.05 & 0.1 & 0.2 & 0.5\\
 \hline
Accuracy & 86.69 & 85.92 & \textbf{89.09} & 88.68 & 87.92\\
\hline
% \Xhline{1.2pt}
\end{tabular*}
\end{table}

\subsubsection{Efforts of Hyper-parameter $\lambda$.}
To validate the effort of hyper-parameter $\lambda$ in eq. (\ref{eq:loss}), we conduct the experiments with different settings on ADNI dataset. As shown in the Table \ref{tab:param}, CvFormer can obtain stable results on the ABIDE dataset in most situations. we can readily find that the proposed TaGBL obtains the best performance when $\lambda=0.1$. More importantly, it's obvious that there exists a wide range for hyper-parameter $\lambda$.

% \subsubsection{Efforts of Hyper-parameters.}
% As demonstrated by the results presented in Table \ref{tab:abalation}, the proposed CvFormer exhibited superior performance when simultaneously utilizing these sub-modules. In comparison to utilizing single RoI or connectivity view, two views could results in more rich information than single view. Despite the use of transformer encoder to combine information between two views, it failed to achieve better performance without cross-view encoder. In terms of fusion strategy, the performances of both the sum and concat strategies were found to be comparable. Consequently, the integration of both the RoI-view and connectivity-view branches could lead to the highest accuracy, indicating that these two branches learn complementary features.

% \begin{table}
% \centering
% % \setlength\tabcolsep{2pt}% 
% \caption{Performance comparison with different $\lambda$ (\%).}
% \setlength{\tabcolsep}{3mm}{
% \begin{tabular}{c|cccccc}
% \hline
% $\lambda$ & 0.001 & 0.005& 0.01 & 0.05& 0.1 & 0.2\\
% \hline
% Accuracy & 80.52 & 82.58 & 80.52 & 82.58 & 85.67 & 82.58 \\
% Recall & 80.52 & 82.58 & 80.52 & 82.58 & 86.72 & 82.58 \\
% \hline
% \end{tabular}
% }
% \end{table}

\section{Conclusion}
This paper proposes a novel transformer model for modeling human brains, called CvFormer. CvFormer is comprised of three main components: tokens block, cross-view transformer block, and fusion block. Unlike previous related works, the token block is capable of producing both RoI-view and connectivity-view tokens, as opposed to just one. The cross-view transformer block, which is the core component of CvFormer, provides two branches based on Transformer modules to process RoI and sub-connectivity tokens. It also implements a cross-view encoder to facilitate the exchange of complementary information between the two views. The final representation of the human brain is obtained through the pooling block, which generates cross-view global tokens. Thereto, CvFormer can simultaneously take into account the diverse and complementary information from both views of the human brain, producing distinct and informative brain representations. Moreover, we propose a two-stage strategy to train its parameters. We combine RoI and connectivity views with contrastive learning to pre-train the CvFormer and fuse cross-view information to finetune the CvFormer using label information. The experimental results can validate the effectiveness of CvFormer.

% \section{Acknowledgements} 
% The authors would like to thank the anonymous reviewers for their insightful comments and suggestions to significantly improve the quality of this paper. 

\bibliographystyle{IEEEtran}
\bibliography{IEEEexample}

\vspace{12pt}

\end{document}